%% file: main.tex
\documentclass[5p, preprint]{elsarticle}
\usepackage[cp1251]{inputenc} 
\usepackage[english]{babel}
\usepackage{amssymb,amsmath}
\usepackage{graphicx}
\usepackage{caption}
\usepackage{subcaption}
\usepackage[colorlinks=true, pdfstartview=FitV, linkcolor=blue, citecolor=blue, urlcolor=blue]{hyperref}
\usepackage{eso-pic,xcolor}
\usepackage{listings} 
\usepackage{multirow}
\usepackage{multicol}
\usepackage{booktabs}
\usepackage{amsthm}
\usepackage{mathtools}
\usepackage{amsmath}
\usepackage{enumitem}
\usepackage{esint}

\usepackage{soul}

\usepackage[noend]{algorithmic}
\usepackage{multicol}
\numberwithin{equation}{subsection}

\journal{Computers and Geotechnics}

\begin{document}

\input{s0_frontmatter.tex}


\input{s1_intro.tex}

\input{s2_overview.tex}
\input{s3_algs.tex}
\input{s4_experiments.tex}

\input{s5_conclusion.tex}
\input{s6_acknowledgements.tex}
\input{s7_references.tex}

\end{document}

%% file: s0_frontmatter.tex
\begin{frontmatter}

\title{
Application of accelerated fixed-point algorithms to hydrodynamic well-fracture coupling
}

\author[smr,mipt]{Vitalii Aksenov}
\ead{aksenov.vv@phystech.edu}
\author[smr]{Maxim Chertov\corref{mycorrespondingauthor}}
\ead{mchertov@slb.com}
\author[smr]{Konstantin Sinkov}
\ead{ksinkov@slb.com}

\cortext[mycorrespondingauthor]{Corresponding author}

\address[smr]{Schlumberger Moscow Research, 13 Pudovkina st., 119285 Moscow, Russia}

\address[mipt]{Moscow Institute of Physics and Technology, 9 Institutskiy per., 141701 Dolgoprudny, Moscow Region, Russia}

\begin{abstract}

The coupled simulations of dynamic interactions between the well, hydraulic fractures and reservoir have significant importance in some areas of petroleum reservoir engineering. Several approaches to the problem of coupling between the numerical models of these parts of the full system have been developed in the industry in past years. One of the possible approaches allowing formulation of the problem as a fixed-point problem is studied in the present work. Accelerated Anderson's and Aitken's fixed-point algorithms are applied to the coupling problem. Accelerated algorithms are compared with traditional Picard iterations on the representative set of test cases including ones remarkably problematic for coupling. Relative performance is measured, and the robustness of the algorithms is tested. Accelerated algorithms enable a significant (up to two orders of magnitude) performance boost in some cases and convergent solutions in the cases where simple Picard iterations fail. Based on the analysis, we provide recommendations for the choice of the particular algorithm and tunable relaxation parameter depending on anticipated complexity of the problem.

\end{abstract}


\end{frontmatter}

%% file: s1_intro.tex
%
%
%
%
%

\section{Introduction}

In this paper, we study the robustness and performance of the algorithms for solution of the fixed-point problem in application to pressure-rate coupling between computational models used in petroleum reservoir simulation. 
Anderson's \cite{AndersonOrig} algorithm and a generalization of Aitken's \cite{aitken1927xxv} algorithm for multidimensional fixed-point problems, referred to as {\it accelerated} algorithms below, are compared to reference Picard iterations~\cite{picard1890memoire}.

Simulation is a common tool to predict and optimize production of hydrocarbons. Often, simulations are focused on either reservoir or well part of the problem, evaluating response of the other part using a simplified steady-state analytical model. This approach may be justified by the huge gap between characteristic times for well and reservoir. However, depending on the nature of production instabilities and transients, temporal and spatial time scales of phenomena occurring in a well and reservoir itself may significantly overlap~\cite{nennie2007investigation}. Accordingly, the coupled simulations taking into account dynamic interaction between well and reservoir and incorporating numerical models for both well and reservoir are needed in certain areas of petroleum reservoir engineering. For the review of particular applications one can refer to \cite{nennie2007investigation, sagen2011dynamic}. 

In~\cite{dasilva2015review} the review of coupled well-reservoir simulations and classification of the possible coupling methods have been presented. Here we follow this classification subdividing the coupling algorithms into {\it fully implicit}, {\it implicit} and {\it explicit}. The {\it fully implicit} approach implies merging of two sets of partial differential equations underlying the numerical simulators into the single set and common discretization. This approach appears to be the most rigorous and is quite commonly used; for example, see~\cite{holmes1998application}. However, it may be impractical from the software development perspective, because it would demand re-implementation of major parts of the simulators to be coupled. In the {\it explicit} approach, the well and reservoir models are solved sequentially, there are no coupling iterations performed within a timestep and essentially there is a mismatch between pressures and rates in the order of one timestep. This approach seems to be rarely used in well-reservoir coupling and will not be further discussed here. In the present work we adopt the {\it implicit} coupling approach that iterates pressure-rate boundary conditions between well and reservoir simulators until they satisfy certain tolerance on each timestep. This approach is described in detail in Section~\ref{sec:overview}. For more examples on applications of this approach, refer to~\cite{nennie2007investigation} and~\cite{bhat2005coupling}. 

The coupling problem in the implicit approach may be cast to the form of the fixed-point problem that can be solved by Picard iterations~\cite{picard1890memoire}. The fixed point iteration method (also known as the Picard algorithm) seems to be often selected when it is necessary to quickly implement a simple algorithm to solve a non-linear matrix equation.
From our past experience~\cite{ChertovChaplygin2019}, a simple Picard algorithm may converge very slowly. It also depends on an arbitrary parameter (relaxation factor), which is selected via trial and error. An unfortunate choice of the relaxation factor slows down the convergence and may often result in divergence of the numerical solution~\cite{ChertovChaplygin2019}. Manual fine tuning of relaxation parameter is very inconvenient, especially when it is necessary to ensure convergence of all cases in a massive parametric study with many simulator runs.

To our knowledge, application of the accelerated algorithms has been never reported in literature on petroleum well-reservoir simulations. On the contrary, the acceleration methods are used within other domains of multiphysics modeling requiring coupling. For instance, one can mention the broad class of fluid-structure interaction (FSI) problems. Fully implicit, implicit and explicit classes of methods mentioned above correspond to monolithic, partitioned and partitioned staggered classes in FSI terminology~\cite{degroote2008stability}. Aitken relaxation found application within the partitioned approach allowing formulation as a fixed-point problem~\cite{mok2001accelerated, AitkenUR}.

The paper is organized as follows. 
Following this introduction,  Section~\ref{sec:overview} gives a brief description of the numerical simulation framework for which the coupling algorithm is needed.
The coupling algorithm is recast mathematically as the fixed-point problem. 
In the numerical framework we consider, the fixed-point problem is complicated by the unavailability of analytical derivatives of the target function and high computational cost to evaluate the target function itself, which scales with the size of a numerical problem. As the coupling solver is an integral part of the larger simulation tool, it would also be time-saving if the method did not require tuning of its parameters for each particular case.
In Section~\ref{sec:algorithms}, we describe advanced Anderson's and Aitken's iterative algorithms and document some practical aspects of their implementation in our in-house code. These algorithms turn out to be the most suitable methods for coupling in our case given the constraints on the computational cost of the target function and unavailability of the derivatives. We also address the theoretically proven convergence properties of these methods and review some examples of previous experience of their usage in numerical modelling. In section~\ref{sec:experiments}, we evaluate in detail the performance of the implemented Anderson's and Aitken's iterative algorithms and compare them against the conventional fixed-point (Picard) iterations. 
The performance of these methods is benchmarked and demonstrated using two numerical problems that are of practical interest for the oil and gas industry and that heavily depend on the well-fracture coupling. 
The first problem is the cascading failure of hydraulic fractures, where failure of one fracture caused by excessively high flow rates can trigger failure of subsequent fractures. The second problem is the severe slugging during the multiphase flow in a horizontal well with multiple fractures. The cascading failure problem is a particularly hard case for the Picard iterations algorithm and highlights 
the benefits of the accelerated algorithms.
The slugging cases evaluate 
robustness and performance of the algorithms for 
a wide range of system parameters and flow regimes.
The detailed study of performance 
provides guidance and recommendations on optimal selection of the control settings of coupling algorithms. Finally, Section~\ref{sec:conclusion} gives a summary of key results obtained in this work.

%% file: s2_overview.tex
\section{Problem overview}\label{sec:overview}
\subsection{General simulator description}
\begin{figure*}[t!]
    \noindent
    \centering
    \includegraphics[width=120mm]{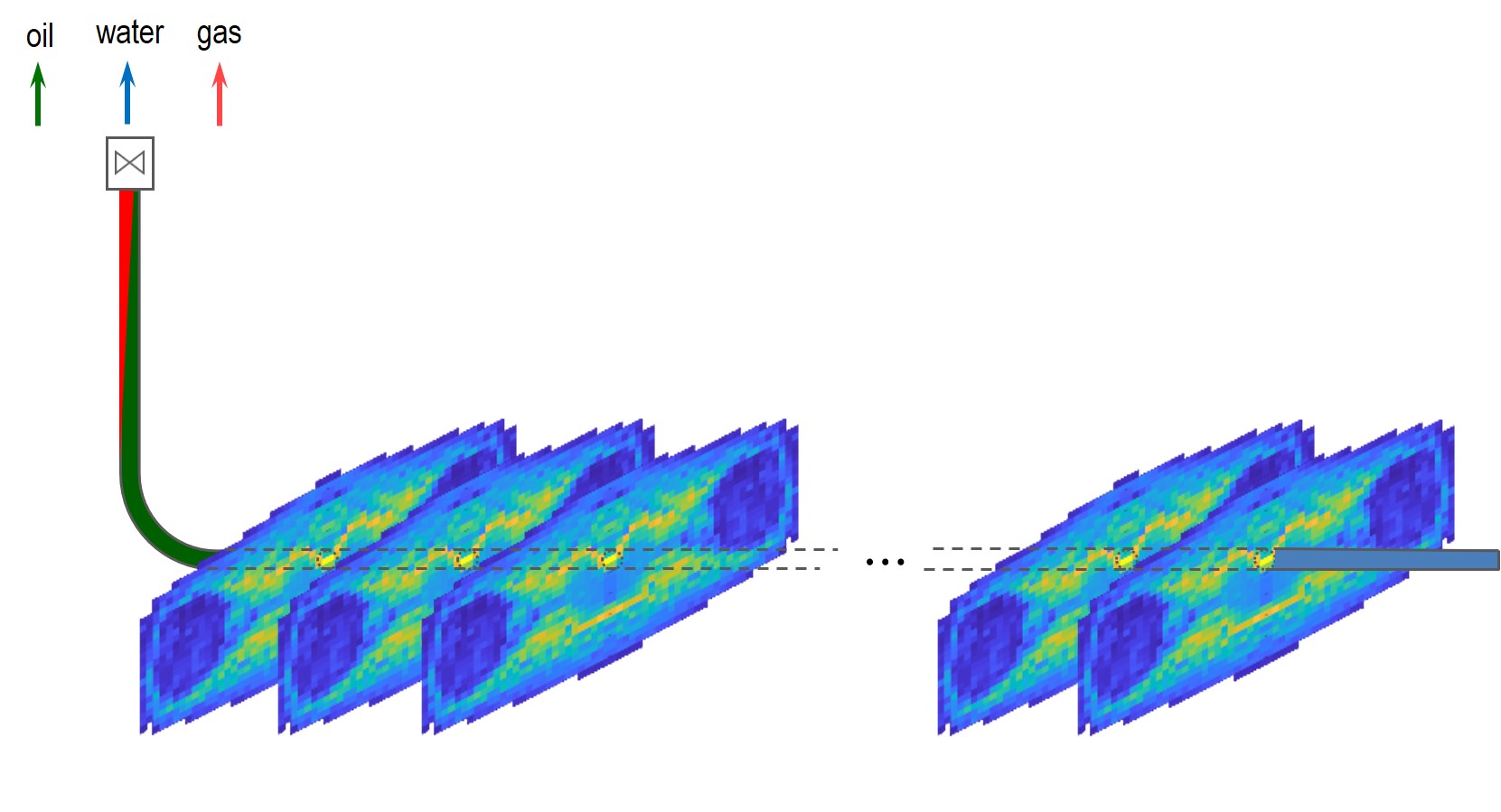}
    \caption{Sketch of the simulated system}
    \label{fig:color_sketch}
\end{figure*}

The problem of interest is the transient multiphase flow in the multistage hydraulically fractured well during well startup and early production after fracturing. The system simulated in the numeric experiments presented below is schematically shown in \figurename~\ref{fig:color_sketch}. It consists of the surface choke (flow restricting device), used to control the well, the well, conducting fluids' flow from subsurface to wellhead, the fractures (increased hydraulic conductivity channels created in rock by high pressure pumping of fluid and filled with propping agent) and the reservoir (layer of porous and permeable rock carrying hydrocarbons). In general, the physical processes taking place in different parts of the system have quite distinct spatial and temporal scales. However, our specific interest to describe the early time behaviour of the well implies that characteristic time scales of the different parts of the system overlap. For example, processes of the fractures and well cleanup are mutually dependent. The degree of displacement of fracturing fluid by reservoir hydrocarbons in the well governs hydrostatic head and therefore the bottomhole pressure. The difference between the bottomhole and reservoir pressures in turn defines production rates from the reservoir and the speed of the displacement.

The problem of the coupled simulation in our case is solved by decomposition into several domains. The dimensions of the different domains vary from a 0D choke model, which is treated as pointwise object and described by the set of algebraic relations, 1D cross-section averaged wellbore flow and reservoir inflow models to a 2D averaged fracture flowback model.

The principal parts of the numerical simulator are the choke module, wellbore flow module, hydraulic fracture module and reservoir module. Below, we will briefly describe physical models underlying each of these modules. Different approaches are used for coupling of different subdomains. In this work, we are focused on the problem of pressure-rate coupling between the well and fractures; accordingly the coupling between in choke-well and fractures-reservoir systems will be only outlined below.

In practice, the choke is used to create additional pressure drop at the wellhead depending on the flow rates of fluids produced. It allows restricting the flow and controling the well. Because of transient nature of the flow characterized by the gradual displacement of the fracturing fluid by liquid and gaseous hydrocarbons, the choke model should take into account presence of multiple phases and effects related to compressibility of gas. Here we adopt the model presented in \cite{alsafran2009predictions}. It is formulated as a system of algebraic equations relating composition of the flow, flow rate, choke orifice diameter and pressures upstream and downstream of the choke. From the computational point of view, the choke module in our simulations calculates wellhead upstream choke pressure given the downstream choke pressure, choke opening, specified as boundary conditions, and flow rates through the choke.

The wellbore flow simulator used in this study has been previously described in \cite{spesivtsev2017study, spesivtsev2013modeling}. The model consists of the 1D cross-section averaged mass and mixture momentum conservation equations. The slip between phases is described according to the drift-flux model \cite{shi2005b}. The wellbore module is governed by the wellhead pressure calculated from choke model and inflow from fractures represented as mass source terms. Given the wellhead pressure and inflows, it returns the surface flow rates and distribution of the bottomhole pressure along the well, in particular, values of pressure in points of connection of well with hydraulic fractures and the wellhead flow rates.
For a given time moment the problem of pressure-rate coupling in choke-wellbore system may be considered as a problem of simultaneous solution of non-linear algebraic equations. The solution is calculated using a Newton-like algorithm for the fixed values of inflows provided by fractures' models.

The numerical model of a hydraulic fracture~\cite{chuprakov2020proppant} is the key component of the simulator and incorporates both hydrodynamical and geomechanical aspects of the problem. Fluid flow relative to the proppant pack inside the fracture is governed by the Darcy law. The fluid phase continuity equation solved is two-dimensional and numerically resolves variation of flow rate inside the fracture. 
The model of fracture closure implemented is local in the sense that it assumes the elastic response of a proppant pack and its resulting width is governed by the local value of effective stress on the pack, mass of proppant per unit area of fracture and compression modulus of the proppant pack. The pack permeability and porosity are also governed~\cite{barree2018generic} by the local effective stress. 

The reservoir matrix model provides the inflow of the flowback fluid into the fracture depending on the difference between pressure inside the fracture and the pore pressure. The model used in the present study is relatively simple and is based on the well-known Carter-type solution~\cite{economides2000}. We calculate the inflow to every discretization cell of the fracture according to the local pressure inside the fracture cell, far field reservoir pressure, porosity and compressibility of matrix and viscosity of the fluid. This approximation corresponds to the assumption of bilinear flow in the reservoir. The assumption is justified by the consideration of the early times after flowback start only. Because the Carter-type inflow is a linear function of pressure it may be implicitly approximated then solving for pressure distribution inside the fracture.

\subsection{Coupling problem}
The need to solve fixed-point nonlinear equations arises from the problem of coupling the rates between well and reservoir numerical models. 
In the well model, the length of the well is discretized into cells $C_i,\ i \in \overline{1, n_{max}}$.
Let $Q_{i,\alpha}$ be the mass rate of sources in $i$-th cell. The index $\alpha$ denotes a rate of a certain phase, e.g., oil, gas, fracturing  fluid, proppant, etc. 
Depending on these rates and the state of the well at a given moment of time $t$, the wellbore flow simulator returns the pressures in $i$-th cell $P_i$. In vector form
$$
    \mathbf{P} = WELL(\mathbf{Q}, t)
$$
The model of reservoir consists of several fractures, each attached to one certain cell of the wellbore. Given pressure $P_i$ in the adjacent cell, the fracture solves flow the problem inside of it and returns the resulting inflow rates. 
$$
    \mathbf{\tilde Q} = FRACTURES(\mathbf{P}, t)
$$
At each timestep the goal is to match the rates $\mathbf{Q}$, used as a boundary condition for the well simulator, to the resulting rates $\mathbf{\tilde Q}$. Thus, we face a fixed-point problem:
\begin{equation}
\label{eq:fixed-point_problem}
    \mathbf{Q} = G(\mathbf{Q}) \left(\ =FRACTURES(WELL(\mathbf{Q}, t), t) \right)
\end{equation}
As solving this equation is an intermediate procedure, done at each time step, the performance of the coupling method is crucial for the simulator's overall performance. 
The method must be sufficiently robust to cope with acute changes of the flow regime, which occur in some cases of practical interest. 
The target function $G$ is acquired via numeric simulation, and the analytical form of its Jacobian is inaccessible, making Newton's method unacceptable.
Due to the high computational costs of evaluating $G$, any schemes including line search or backtracking~\cite[Chapter~9.7]{press2007numerical} are also inappropriate. 
Methods based on approximation of the Jacobian or its inverse with full rank matrices, 
    such as Broyden methods, are not desirable due to the scale of the problem. 
Indeed, the dimension of variables vector $\mathbf{Q}$ is $n_\alpha \cdot n_c$. 
Here $n_\alpha = 4$ is the number of phases of the flow (oil, water, gas, proppant), 
    and $n_c$, which is the number of cells that have a fracture attached to them, may be as large as $n_c \sim 100$, as in one of the cases studied below. 
This yields a Jacobian with $\sim 10^4 \div 10^5$ components. 
Storage of such a large matrix is too demanding for our purposes. 
Additionally, for such methods, an initial guess for inverse Jacobian is required, and constructing such a guess for each timestep in the absence of an actual analytical Jacobian is a nontrivial problem itself.
Thus we require an acceleration algorithm  which
\begin{itemize}
    \item Stores only a fixed small amount of previous iterates that do not depend on the dimension of $\mathbf{Q}$
    \item Does not evaluate $\mathbf{Q}$ too often, e.g. for backtracking
    \item Does not depend on more than one arbitrary parameter (like the relaxation factor in Picard method)
\end{itemize}

%% file: s3_algs.tex
\section{Algorithms}\label{sec:algorithms}

From here and further, $\left\{x_i\right\}_{i=0}^{\infty} $ denotes a sequence of points, generated by some iterative process, 
    $x^*~:~x^*~=~G(x^*)$ is the solution to the fixed-point problem;
    $g_i~=~G(x_i)$; $f_i~=~g_i~-~x_i$ is the residual vector. 
    If not explicitly said otherwise, $\langle \cdot , \cdot \rangle$ represents scalar product, $\| \cdot \|$ is the Euclidean norm in $\mathbb{R}^n$. 
    \subsection{Fixed-point iteration with relaxation (Picard method)}
        \hrule
        \begin{algorithmic}[1]
            \REQUIRE $x_0$~---~initial guess, $\beta \in (0;1]$~---~relaxation factor
            \REPEAT
            \STATE $x_{k + 1} = \beta g_k + (1 - \beta)x_k$
            \UNTIL{checkStopCondition() \OR $k \geq k_{max}$}
            \end{algorithmic}
        \hrule
        \medskip
        This is the most straightforward solution. 
        The user-defined parameter $\beta$ determines the length of the algorithm step. 
        The intuition behind its choice is that using $\beta$ close to zero leads to better robustness of the algorithm at the cost of convergence speed. 
        This intuition is supported by strict proof when $G$ is a \emph{contraction mapping}, or, at least \emph{non-expansive mapping}, 
        see \cite{RyuBoydPrimer}.
        In the former case, the convergence is linear with optimal choice of $\beta = 1$ and convergence speed estimate~is
        $$
            \|G(x_k) - x_k\| \leq c^k\|G(x_0) - x_0\|
        $$
        where $c$ is the contraction factor of $G$.
        In the latter case, estimation for the residual is
        $$
            \min_{j\in\overline{0, k}}\|G(x_j) - x_j\| \leq \frac{\|x_0 - x^*\|}{(k + 1)\beta(1-\beta)}
        $$
        with optimal value $\beta = \frac12$. The latter asymptotic is worse that the former, but the subset of $\mathbf{R}^n$ on which the map is non-expansive might be larger than the subset where it is a contraction. 
        Thus, usage of relaxation factor $\beta \neq 1$ contributes to robustness but decreases speed of convergence.
        
    \subsection{Aitken relaxation}
        The first accelerated method we consider in this paper is \emph{Aitken relaxation}. 
        The idea of the method is to change the relaxation factor (and, thus, the size of the iteration step) based on the information from the previous iteration. 
        The update  method for the vector case was suggested in \cite{IronsTuck1969}.
        We used it in a slightly different form, adapted from \cite{AitkenUR}:
        \medskip
        \hrule
        \begin{algorithmic}[1]
            \REQUIRE $x_0$~---~initial guess, $\beta_0 \in (0;1]$~---~initial relaxation factor
            \REPEAT
            \STATE \begin{equation}\label{eq:ATK_rf}
                \beta_k = -\beta_{k-1}\frac{\langle f_{k - 1}, f_{k} - f_{k - 1}\rangle}{\langle f_{k} - f_{k - 1}, f_{k} - f_{k - 1}\rangle}
            \end{equation}
            \COMMENT{Acquire new relaxation factor}
            \STATE $x_{k + 1} = \beta_k g_k + (1 - \beta_k)x_k$
            \UNTIL{checkStopCondition() \OR $k \geq k_{max}$}
            \end{algorithmic}
        \hrule
        \medskip
        In \cite{AitkenLin} the authors consider the case where target function can be written as
        $$
            G(x) - x^* = \mathbf{Q}(x - x^*) + O(\|x - x^*\|^2)
        $$
        where $\mathbf{Q}$ is a Hermitian matrix. The condition for acceleration of the convergence is that all the eigenvalues of the matrix $\left( \mathbf{Q - E} \right)$ have the same sign, where $\mathbf{E}$ is identity matrix.
        
        Consider that the last value of relaxation parameter $\beta_n$ at timestep $t_k$ can be used as initial value $\beta_0$ at timestep $t_{k + 1}$. 
        There is a possibility that such approach can help improve overall performance of the simulator, which is studied below.
        
    \subsection{Anderson's acceleration}
        \hrule
        \smallskip
        \begin{algorithmic}[1]
            \REQUIRE $x_0$~---~initial guess, $\beta \in (0;1]$~---~relaxation factor, $m \in \mathbb{N}$
            \REPEAT
            \STATE $m_k = \min{\{k, m\}}$
            \STATE \begin{equation}\label{eq:AAopt}
                         \alpha = \arg\min_{\alpha}{\left\|\sum_{i =    k - m_k}^k \alpha_i f_i\right\|}\ s.t. \
                            \sum_{i = k - m_k}^k\alpha_i = 1
                    \end{equation}
            \STATE \begin{equation}\label{eq:AAupd}
                         x_{k + 1} = \sum_{i = k - m_k}^k \alpha_i 
                                \left(
                                    \beta g_i + (1 - \beta) x_i 
                                \right)
                    \end{equation}
            \UNTIL{checkStopCondition() \OR $k \geq k_{max}$}
            \end{algorithmic}
        \hrule
        \medskip
        The rationale under this algorithm is as follows. Consider $G$ to be linear: $g_i = \mathbf{G}x_i$.
        Then 
        \begin{gather}
            \arg\min_{\alpha}{\left\|\sum_{i = k - m_k}^k \alpha_i f_i\right\|} = %
                \arg\min_{\alpha}\left\|{\mathbf{(G - I)}\sum_{i = k - m_k}^k \alpha_i x_i}\right\| \notag \\
            \Bar{x} = \sum_{i = k - m_k}^k \alpha_i x_i \notag \\
            x_{k+1} = \beta\sum_{i = k - m_k}^k \alpha_i \mathbf{G}x_i + (1 - \beta)\sum_{i = k - m_k}^k \alpha_i x_i = \notag \\
                = \beta \mathbf{G}\Bar{x} + (1 - \beta)\Bar{x}
        \end{gather}
        i.e., Picard step is done from $\Bar{x}$~---~the point that minimizes residual over the affine hull of $m_k$ previous iterates. 
        In \cite{WalkerPengAnderson} the authors prove general equivalence between Anderson's acceleration and GMRES.
        In \cite{Fang2009TwoCO} Anderson acceleration is considered as a member of a broader family  of multisecant methods (in particular, Broyden's family Type-II method).
        
        The existing local convergence theory for general nonlinear case, to the authors' knowledge, is not promising. 
        Consider, for example, the sufficient conditions of convergence for Anderson acceleration with $m = 2$, proposed in \cite{TothKelleyConvergence}. 
        When $G$ is a contraction, one of the requirements is that contraction factor $c$ is sufficiently small, so that $\hat c = \frac{3c - c^2}{1 - c} < 1$.
        Then the algorithm converges linearly with convergence factor not greater than $\hat c$: 
        $$
            \|x_k - x^*\| \leq C{\hat c}^k
        $$
        In the same conditions, Picard method would converge with factor $c$.
        If the value of $c$ is small, $\hat{c} \approx 3c$, so the guaranteed convergence rate for Anderson acceleration is worse. 
        On the other hand, it is reported (see \cite{WalkerPengAnderson, LipnikovSvyatskiyVassilevski}) that Anderson acceleration actually improves convergence of the coupled iterations. 
        A comprehensive study of application of Anderson's acceleration to the problem of variably saturated flow is presented in \cite{LottWalkerWoodwardYang}.
        The computational experiments, conducted by the authors, demonstrate greater robustness of the algorithm compared to Picard accelerations. 
        
        Note that the choice of relaxation parameter $\beta$ is solely up to the researcher.
        In the works \cite{LipnikovSvyatskiyVassilevski, LottWalkerWoodwardYang} the authors formulate the Anderson acceleration algorithm without introducing this parameter, 
            in a way which is in our notation equivalent to substituting $\beta \equiv 1$ in formula \ref{eq:AAupd}.
        In \cite{WalkerPengAnderson, TothKelleyConvergence, AndersonOrig} both theoretical and numerical findings are provided for the case $\beta = 1$.
        In \cite{Fang2009TwoCO} successful computations with $\beta$ other than $1$ are reported, but, unfortunately, no rationale under the particular choice is provided.
        In the following numeric experiments we will try to address the problem of dependence of convergence speed of Anderson's method on choice of $\beta$.
        
        \subsection{Implementation details} 
        In our implementation of Anderson's algorithm, we use only information from two previous points ($m=2$). 
        The optimization problem \eqref{eq:AAopt} can then be easily solved analytically:
        \begin{gather}
            \alpha_k = \alpha,\ \alpha_{k-1} = 1 - \alpha \notag \\
            \alpha = \arg\min_{\alpha \in \mathbb{R}}{\|\alpha f_k + (1 - \alpha)f_{k-1}\|} \notag \\%
            \alpha = \frac{\langle f_{k-1} - f_k, f_{k-1} \rangle}{\langle f_{k-1} - f_k, f_{k-1} - f_k \rangle}\label{eq:TPAopt}
        \end{gather}
      
        Computational~formulas~for~both~~Aitken's~and~Anderson's~~algorithms~~contain~a~term $\langle f_{k-1}~-~f_k,~f_{k-1}~-~f_k \rangle$ in the denominator (see \eqref{eq:ATK_rf}, \eqref{eq:TPAopt}).
        When two consecutive residuals are close to being equal, the mentioned denominator becomes small. 
        In our practice, this resulted in unreasonably large iteration steps, causing oscillation of iterates, which slows down the convergence. 
        Thus, in the numeric implementation, we apply the constraint
        \begin{equation*}
        \|f_{k}~-~f_{k-1}\|~\geq~\epsilon~=~10^{-10} 
        \end{equation*}
        In case that constraint is not satisfied at the point $x_k$, a Picard step with relaxation factor $\beta_k$ or $\beta$ for Aitken's and Anderson's methods, respectively is done .
        The stop condition in all the algorithms~is 
        \begin{equation}\label{eq:stop_condition}
            \|g_k - x_k\| \leq \varepsilon \|x_k\|
        \end{equation}
        where $\varepsilon$ is the required relative tolerance.

%% file: s4_experiments.tex
\section{Numeric experiments}\label{sec:experiments}

In the series of numerical experiments considered in this section we assess the performance of the implemented accelerated algorithms (Anderson's and Aitken's) compared to the widely used conventional fixed-point iterations.
For brevity, they are abbreviated as TPA, \emph{two-point Anderson's acceleration}; ATK, \emph{Aitken's relaxation} and FPI, \emph{fixed-point iteration (Picard's method)}. 

The fixed point problem is formulated in terms of mass rates of each phase normalized by standard densities. 
Rates were chosen instead of pressure because, although rate error scales with pressure error, the scaling coefficient is dependent on the physical dimensions and properties of multiple system components, 
    which is difficult to estimate beforehand. 
This scaling may also be highly variable during the simulation. At the same time, formulation in terms of rates ensures that mass balance conservation is directly controlled by the user-selected convergence tolerance of the iterative algorithm.

The algorithms were run with the same relative tolerance $\varepsilon$. 
We also impose a constraint on the maximum number of iterations allowed at a physical timestep.
If the number of iterations exceeds some predefined value $N_{crit}$, which is the same for all three solvers, the whole simulation is considered diverged.
Under this requirement, all solvers generated the same numerical solution within controlled tolerance consuming different numbers of iterations.
The tests are run on the same predefined sequence of timesteps for each of the solvers. 

The main goal of the study is to establish recommendations on the usage of these methods and provide guidelines for selection of suitable settings of the numerical algorithms. 
In particular, it would be useful to establish some rationale for the choice of relaxation factor $\beta$ ($\beta_0$ in ATK) in each of the methods. 
For that, we perform extensive tests with different relaxation factors for each solver.

The performance of the algorithms is evaluated and benchmarked using two numerical problems involving hydrodynamic coupling between the horizontal well and multiple hydraulic fractures, inspired by oilfield applications.
The first problem is cascading failure of hydraulic fractures due to aggressive flowback. 
The second problem is severe slugging during multiphase flow in horizontal well. 
The cascading failure problem involves rapid changes of flow rates caused by choke adjustment and fracture failure and very well highlights benefits of accelerated algorithm compared to conventional FPI. 
The slugging case evaluates behavior of iterative algorithms in a wide range of physical system parameters. 

In the coupling problem, the time of the individual iteration is defined by the time of evaluation of the $G$ function. 
In our case, it is the total computational time of fracture and wellbore flow simulators.
For all three algorithms, this time is much larger than all the additional operations (vector products and summations) needed to generate the next approximation.
Thus, the  total amount of iterations throughout the simulation is an adequate measure of the overall algorithm performance. 

\subsection{Cascading failure of fractures}
\begin{figure*}[ht]
        \noindent
        \centering
        \includegraphics[width=120mm]{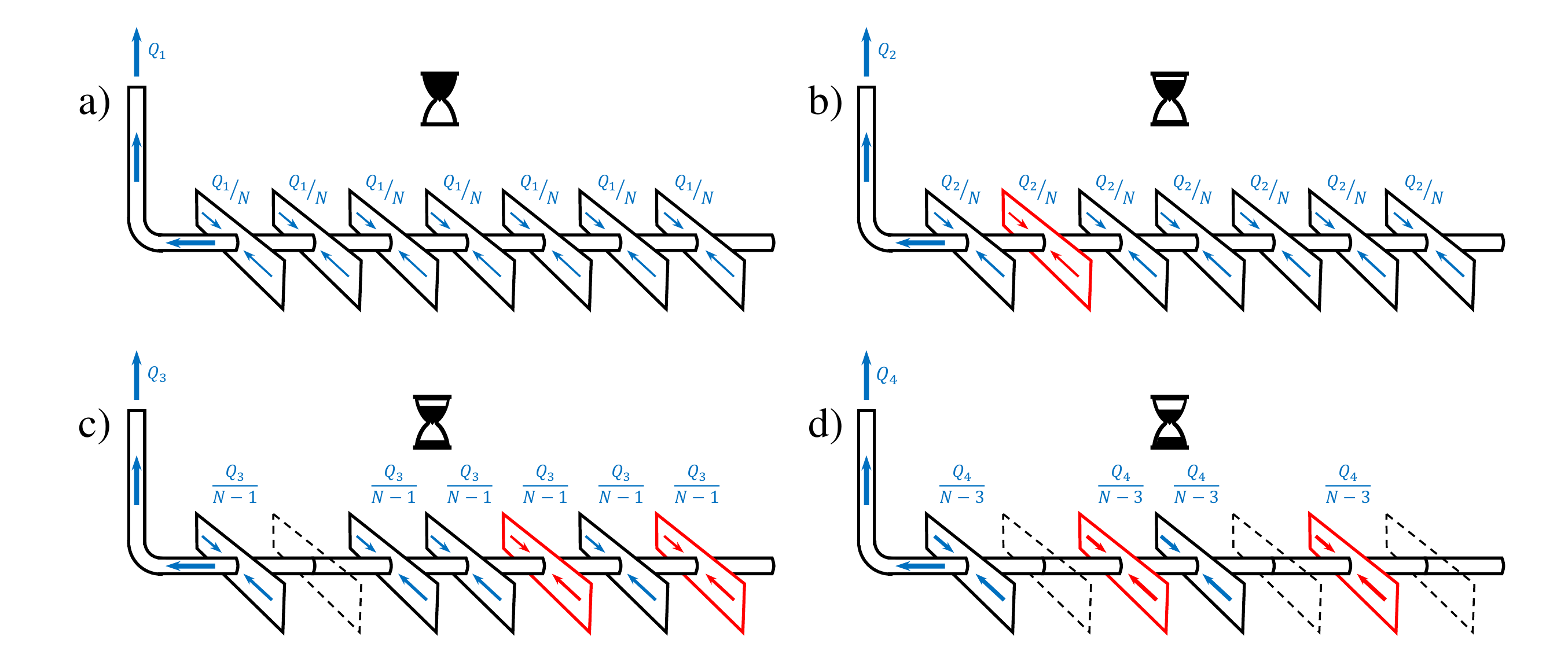}
        \caption{Sequential failure of hydraulic fractures}
        \label{fig:cf_sketch}
    \end{figure*}

One of the undesired phenomena that sometimes takes place in the oilfield industry is the damage of recently created hydraulic fracture conductivity due to aggressive flowback or excessively high flow rate during initial production. High production rates can cause mobilization of proppant in the near-wellbore zone of the hydraulic fractures experiencing the largest pressure gradients and fluid drag acting on the proppant pack. Washout of proppant from the near-wellbore region can result in drop of fracture conductivity and decrease overall fracture productivity. In the most critical cases, this can lead to formation of a pinch-out zone inside the fracture and disconnect it from the well. During flowback in multistage fractured wells, loss of productivity of one of the fractures can trigger the failure of others. In a system of multiple fractures producing into the same well with a single rate control by the surface choke, the failure of one producing fracture redistributes the full production rate between remaining ones. This happens because as the total flow rate decreases, due to the presence of choke, the bottomhole pressure decreases, and then, in turn, the drawdown and the rate from other fractures increases. As discussed in~\cite{chertov2020}, in case of a large number of producing fractures the total rate produced from the well after redistribution is almost unchanged. The redistribution leads to the increase of production rate from the remaining fractures, a subsequent increase of the drag forces on proppant packs inside them and brings remaining fractures closer to failure. The sequence of hydraulic fracture failures is schematically shown in \figurename~\ref{fig:cf_sketch}. Accordingly, the failure of hydraulic fractures in multistage wells due to aggressive flowback is a process with positive feedback. If multiple fractures are close to the critical proppant pack stability condition, the failure of the weakest fracture can trigger the cascading failure of many fractures. This phenomenon, named \emph{cascading failure of fractures}, was introduced in \cite{CascadingFailure} and studied numerically for different criteria of individual fracture failure in~\cite{chertov2020}. 

Note, that in simulations here we consider an idealized model of fracture failure. It is assumed that fractures immediately and completely lose their conductivity if the fluid velocity in proppant pack exceeds some threshold value. As it was demonstrated in~\cite{chertov2020}, this idealized model of fracture failure tend to generate more steep and dramatic change in overall well productivity compared to explicit modeling of proppant wash-out from fracture and associated gradual degradation of their productivity. The preference on the abrupt fracture failure is justified by the goal of testing and evaluation of coupling algorithms in the most troublesome statement. In terms of the fixed-point problem, it means that that the solution from the previous time step used as the initial guess while solving~\eqref{eq:fixed-point_problem} can be a poor approximation for the solution at the next time step and behaviour of the $G$ function will change significantly form one time step to another. 

\begin{figure*}[ht]
        \noindent
        \centering
        \includegraphics[width=120mm]{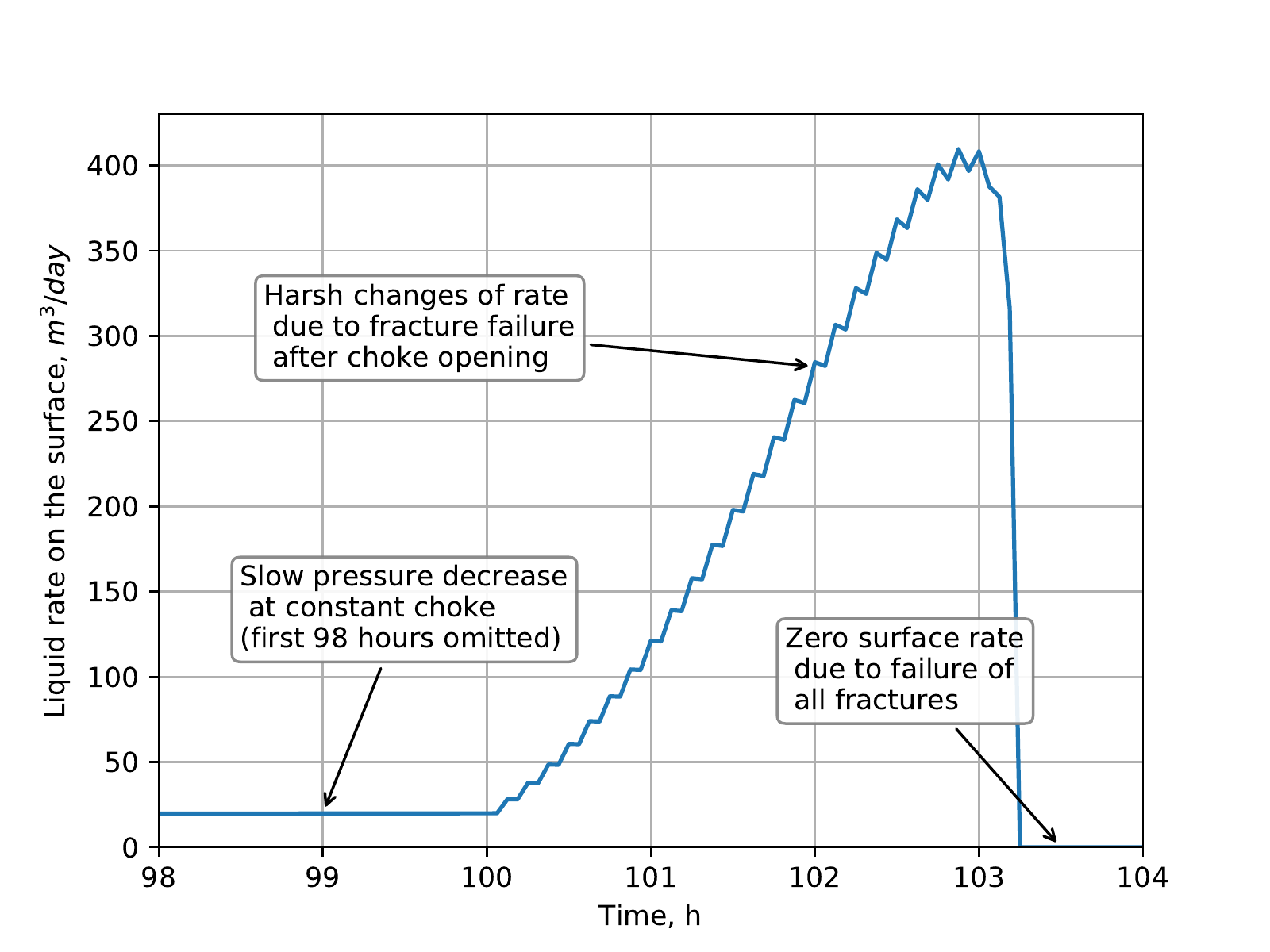}
        \caption{Surface rate for the cascading failure test case}
        \label{fig:cf_demo}
    \end{figure*}

    \figurename~\ref{fig:cf_demo} demonstrates the total production rate from the multistage horizontal well with $100$ hydraulic fractures calculated during numerical simulation of the cascading failure case. To give a flavour of the physical properties of the system, the well is composed of $3000$ m long vertical section and $2020$ m long horizontal section. There are $100$ hydraulic fractures connected to horizontal section with equal spacing of about $20$ m between them. Each fracture is connected to the well by a single cell. Once the critical fluid velocity in this cell is reached, the cell is set nonconductive and the entire fracture is disconnected from the well. During the first $100$ hours, the choke diameter $c = 4/64''$ is kept constant, and wellhead downstream choke pressure $p_{whdc}$ is gradually reduced about twice, from $195.7$~bar to $100$~bar. After that, $p_{whdc}$ is kept constant, and choke opening diameter is stepwise increased by $c = 0.75/64''$ every $450$s. After each stepwise increase of the choke, the total flow rate is stepwise increased, as shown in \figurename~\ref{fig:cf_demo}. Critical velocity to disconnect each fracture is linearly varied within certain range. Initially, flow rate increase after each choke increase is insufficient to fail any fractures. Then, as the number of producing fractures is still large, each choke increase may fail a few fractures and may also potentially trigger cascading failure of a few more fractures due to rate redistribution. At this stage, the total flow rate is nearly flat between stepwise choke openings. As the number of producing fractures goes down, redistributed increase of the flow rate on remaining fractures goes up, the number of sequentially triggered rate redistributions and the number of fractures dropped after each redistribution goes up. As a result, each stepwise increase of the flow rate is now followed by a decreasing slope of flow rate. In the end, the last choke opening initiates the last  cascade of fracture failures, when there is no combination of fractures that can support the flow rate, and all remaining fractures fail in a sequence.  
   
Because FPI is the simplest and most widely applied method, we will estimate the relative acceleration provided by TPA and ATK as $S = \frac{N(\text{FPI})}{N(TPA)}$, or $\frac{N(\text{FPI})}{N(ATK)}$, where $N()$ is the total number of iterations for each algorithm. \tablename~\ref{tab:cf} gives the summary of the benchmarking on the cascading failure case at different values of relaxation factor parameter. $N_{opt}$ is the number of iterations consumed in the best case, which occurred at relaxation parameter $\beta_{opt}$. Relative acceleration $S$ in \tablename~\ref{tab:cf} is presented for the best run of each algorithm.

    \begin{table*}[ht]
        \centering
        \input{tab/cascading.tex}
        \caption{Results of numerical experiments for Cascading Failure case}
        \label{tab:cf}
    \end{table*}
    For all the solvers the required relative tolerance $\varepsilon = 10^{-8}$ and maximal iterations per timestep $N_{crit} = 10^4$. 
    In these conditions, the FPI solver managed to converge only with rather small relaxation factor $\beta = 10^{-3}$. 
    The total amount of iterations with FPI was over $600,000$. \
    The use of Anderson's method with $\beta = 1$ allowed acquiring drastic acceleration of around $600$ times, though even for the worst choice of $\beta = 10^{-3}$ the speedup is around~$12$.
    The findings also demonstrate a complicated nonmonotonous dependence of the performance of the TPA solver on the relaxation factor. 
    The general trend that applies for $\beta$ close both to $0$ and to $1$ is that TPA converges faster for larger relaxation factors. 
    On the other hand, some peculiarities occur in the middle of the range. 
    For example, we can see that for $\beta = 0.5$ the solver works more than~$5$ times faster than for relatively close values $0.3$ and~$0.7$.
    This suggests that some tuning of $\beta$ may be needed to achieve maximal performance at any given case.
    
    For the ATK solver, acceleration is around $11$ times for the best $\beta$ and $9$ for the worst. 
    Though not as dramatic as for TPA solver, this also makes the total computational time acceptable for practical studies. 
    It is also interesting that the total amount of iterations is weakly dependent on initial values of $\beta$, unlike Anderson's solver.
    The number of iterations for different relaxation factors varies within approximately $10\%$ of average. 
    This is because in our implementation of Aitken solver, we take $\beta_0$ on the next timestep equal to $\beta_l$~---~the last value on previous timestep.
    With more iterations, $\beta_k$ less and less depends on the initial choice of $\beta$.
    For practical use, that means that ATK provides good acceleration over FPI for all initial values of relaxation parameters and fine tuning of initial $\beta$ is not necessary.

\subsection{Severe slugging}
\begin{figure*}[ht]
    \centering
    \begin{subfigure}[t]{0.185\textwidth}
    \centering
    \includegraphics{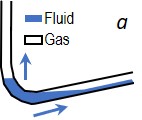}
    \phantomcaption{}
    \label{fig:slugging_stages_1}
    \end{subfigure}
    \begin{subfigure}[t]{0.185\textwidth}
    \centering
    \includegraphics{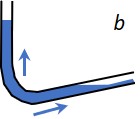}
    \phantomcaption{}
    \label{fig:slugging_stages_2}
    \end{subfigure}
    \begin{subfigure}[t]{0.185\textwidth}
    \centering
    \includegraphics{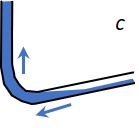}
    \phantomcaption{}
    \label{fig:slugging_stages_3}
    \end{subfigure}
    \begin{subfigure}[t]{0.185\textwidth}
    \centering
    \includegraphics{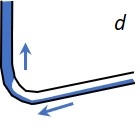}
    \phantomcaption{}
    \label{fig:slugging_stages_4}
    \end{subfigure}
    \begin{subfigure}[t]{0.185\textwidth}
    \centering
    \includegraphics{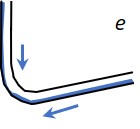}
    \phantomcaption{}
    \label{fig:slugging_stages_5}
    \end{subfigure}
    \caption{Stages of severe slugging}
    \label{fig:slugging_stages}
\end{figure*}

Slugging flow is a special flow regime possible in pipelines and wellbores transporting gas and liquid simultaneously. Slugging flow is characterized by unstable system behavior with significant variations of pressure and flow rates in space and time. These variations can be observed even for steady-state boundary conditions such as constant pressure at the pipeline segment outlet and constant gas and liquid rates at the pipeline inlet. In hilly terrain pipelines, the undulating trajectory may lead to the formation of terrain-induced slugs (for example,~\cite{taitel1990,dehenau1995,spesivtsev2017study}). If the configuration consists of a near-horizontal pipeline or a well connected with a vertical riser (\figurename~\ref{fig:slugging_stages}), the transient behavior observed in such systems is referred to as severe slugging (for example,~\cite{fabreetal1990,balino2010}). The physical mechanisms leading to the formation of the slug flow are the same in both cases. The effect is caused by the competition between gravity-dominated liquid flow and compressibility of the gas phase. The liquid might accumulate in the lower parts of the system and subsequently block the free gas flow (see~\figurename~\ref{fig:slugging_stages_1}). The compressibility of gas and continuous inflow to the pipe allows the pressure to build up gradually in the trapped gas volume (\figurename~\ref{fig:slugging_stages_2},~\ref{fig:slugging_stages_3}). Once the gas pressure reaches some threshold, rapid outflow from the system occurs (\figurename~\ref{fig:slugging_stages_4}) and the process is repeated (\figurename~\ref{fig:slugging_stages_5}). The terrain-induced slugging is generally treated as an undesirable phenomenon, as it may result in the separator overflow and damage of the surface equipment.
    
    \begin{figure*}[ht]
        \noindent\centering{
        \includegraphics[width=150mm]{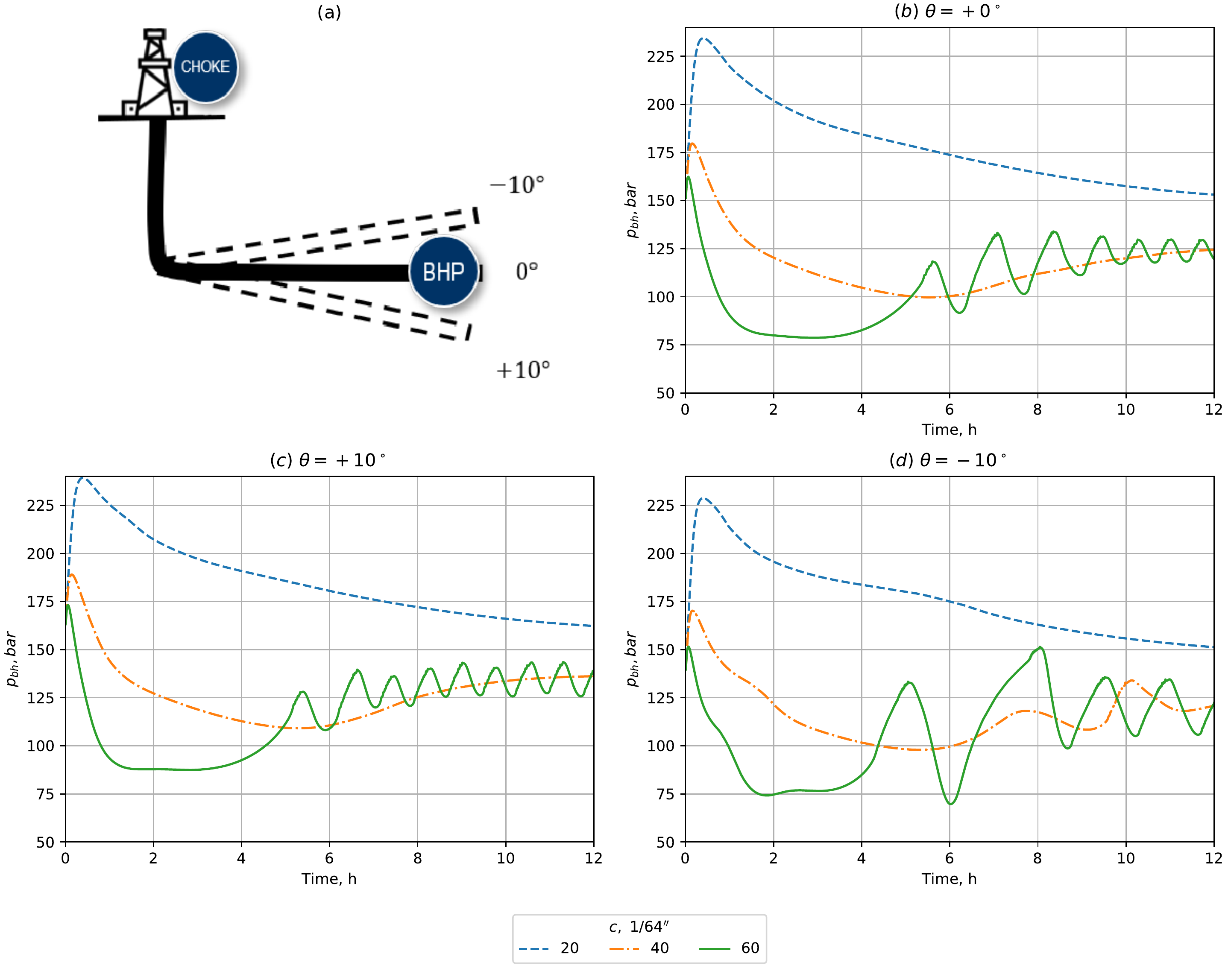}
        \caption{Toe-up, horizontal and toe-down wells with bottomhole pressure behavior for different choke diameters $c$}}\label{fig:slug_demo}
    \end{figure*}
    
    For the parametric study of the process we have designed a simplified model of the well in which slugging can occur. 
    It consists of a vertical and an inclined sections, with lengths $L_{v} = 2$ km and$\ L_i = 1$ km respectively. 
    The inclination of the horizontal section is characterized by the angle $\theta$;  $\theta > 0$ represents the toe-down well and $\theta < 0$~---~toe-up. $N_{fracs} = 20$ fractures are located on the inclined section with equal spacing. 
    This simple setup geometry is sufficient to develop a slugging flow regime.

    From the computational point of view, this problem was chosen because it can provide a set of test cases which can be defined by a small amount of varying parameters (in particular, choke size and the inclination angle $\theta$), but provide qualitatively different flow regimes (oscillatory when slugging occurs, relaxation to steady-state otherwise).
    
    From our previous experience, we anticipated that there were two main factors that may inhibit the convergence in this parametric study.
    First, we expected fast changes in the solution when the flow regime is oscillatory.
    When such changes occur, the initial approximation taken from the previous step may be far from the solution on the next step, so the convergence might slow down.
    Second, we expected that the behavior of the target function $G$ can vary with physical parameters. In particular, when the choke diameter is small, the dependence of pressure on rate becomes steeper. Although the analytical form of $G$ is not available, this suggests that convergence at small chokes can be hindered.

    The goal of the numerical experiments with slugging is to asses the performance of the coupling algorithms on a set of cases with different flow regimes and properties of target function.
    The dependence of the performance on the choice of  relaxation factor is also studied to provide further insights on its role in convergence.
    
    The set of cases is generated by variation of the inclination $\theta \in \{-10^\circ, -5^\circ, 0^\circ, 5^\circ, 10^\circ\}$ and the choke opening $c \in \{15/64'', 20/64'', 30/64'', 40/64'', 50/64'', 60/64''\}$. The choke opening is kept constant throughout the entire simulation that covers $T = 12$ hours. To investigate the dependence of the solvers' performance on the time step, 
        we performed a parametric study for two time steps: $\tau = 150$ s and $\tau_{small} = 20$ s. The relative tolerance was $\varepsilon = 10^{-13}$ and maximal number of iterations per timestep $N_{crit} = 1000$.

    \begin{table*}[ht]
        \noindent
        \centering
        \input{tab/slug150_var.tex}
        \caption{Results of parametric study for time step $\tau = 150$ s}
        \label{tab:slug_dt150}
    \end{table*}

    \begin{figure*}[ht]
        \noindent
        \centering
        \includegraphics[width=\textwidth]{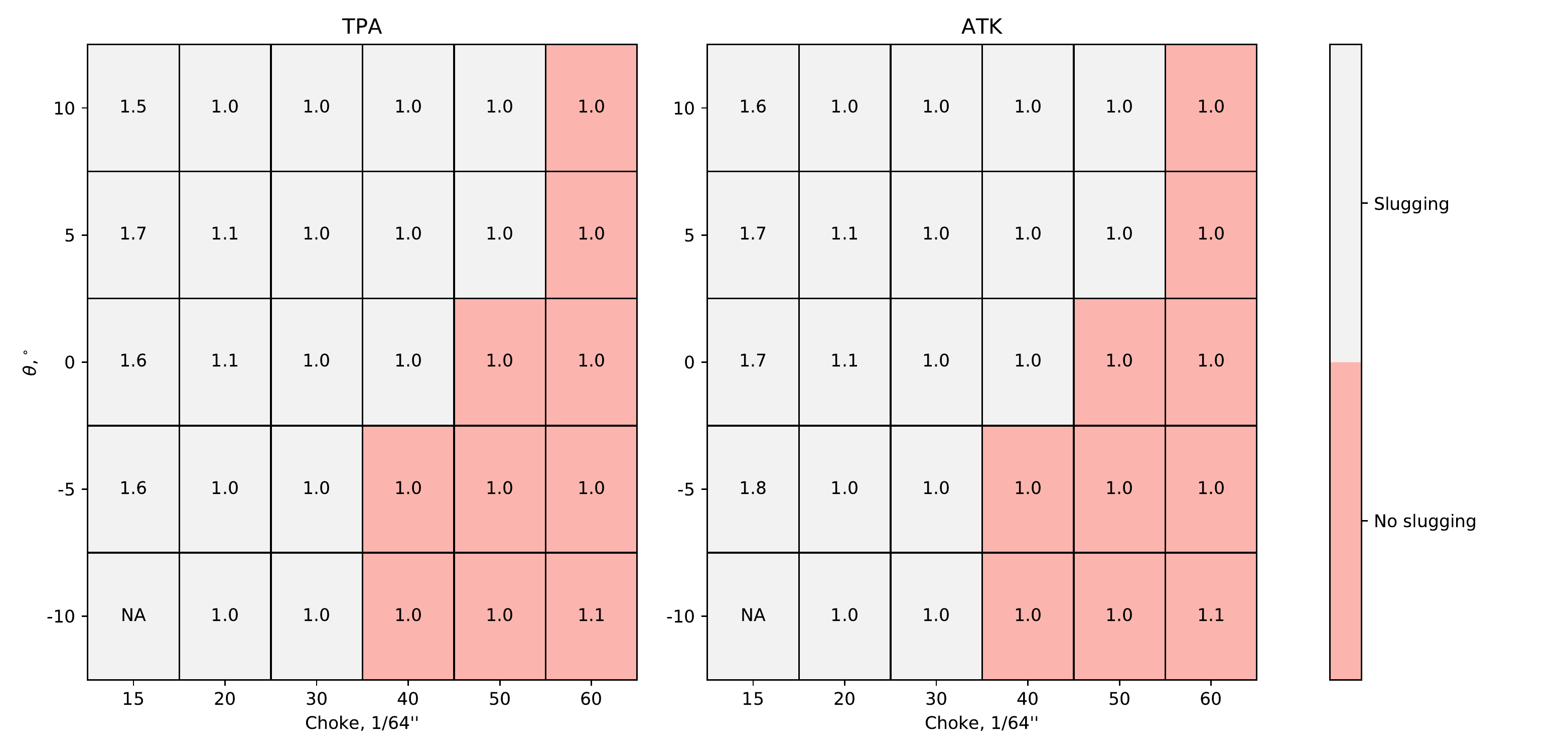}
        \caption{Maximal acceleration for  $\tau = 150$ s }
        \label{fig:acceleration150}
    \end{figure*}

    A representative slice of results of the test runs for time step $\tau = 150$s is compiled in \tablename~\ref{tab:slug_dt150}. 
    The maximum acceleration for TPA and ATK relative to FPI and the map of slugging/non-slugging flow regimes in $(c,\theta)$ coordinates is depicted in \figurename~\ref{fig:acceleration150}.
    For the cases with small choke diameter $c~\in~\{15/64'', 20/64'', 30/64''\}$, the slugging did not occur for any inclination angles. For $c \in \{40/64'', 50/64''\}$, it appeared only for toe-up and horizontal wells ($\theta \leq 0$). For the cases with wide open choke ($c = 60/64''$), oscillations were present  at all inclinations. These observations are in qualitative agreement with generally accepted conclusion that choking effectively eliminates or reduces severity of slugging~\cite{jansen1996elimination}. 

    One notable finding is that the total amount of iterations does not vary from non-oscillatory to oscillatory flow regime with same choke value as much as it varies from smaller choke values to bigger ones.
    Roughly speaking, all cases with small chokes are harder for all three iterative solvers than all cases with large choke. For a set of cases with fixed choke, oscillatory cases, if present, are only slightly harder than non-oscillatory ones. 
    For the case with the smallest choke and steepest toe-up well all three solvers diverged (marked N/A on \figurename~\ref{fig:acceleration150}).
    The maximum acceleration ($1.8$ for ATK and $1.6$ for TPA) is achieved in cases with the smallest choke value.     
    For larger chokes, this acceleration is $[0.98, 1.1]$, so it is fair to say that acceleration is achieved only for a limited set of modeling conditions. 
    Based on these data, we can assume that the factor of steeper change of target function $G$ is more detrimental for convergence than the factor of generating a bad initial approximation due to oscillations.
    The former factor only comes to play for the case $c = 60/64'',\ \theta = -10^\circ$.
    All algorithms require more iterations on this case, and accelerated solvers start to demonstrate minor acceleration of about $1.1$ compared to almost no acceleration in nearby cases with large choke.

    \begin{table*}[ht]
        \centering
        \input{tab/slug20_var.tex}
        \caption{Results of parametric study for time step $\tau_{small} = 20$ s}
        \label{tab:slug_dt20}
    \end{table*} 
    
    \begin{figure*}[ht]
        \noindent
        \centering
        \includegraphics[width=\textwidth]{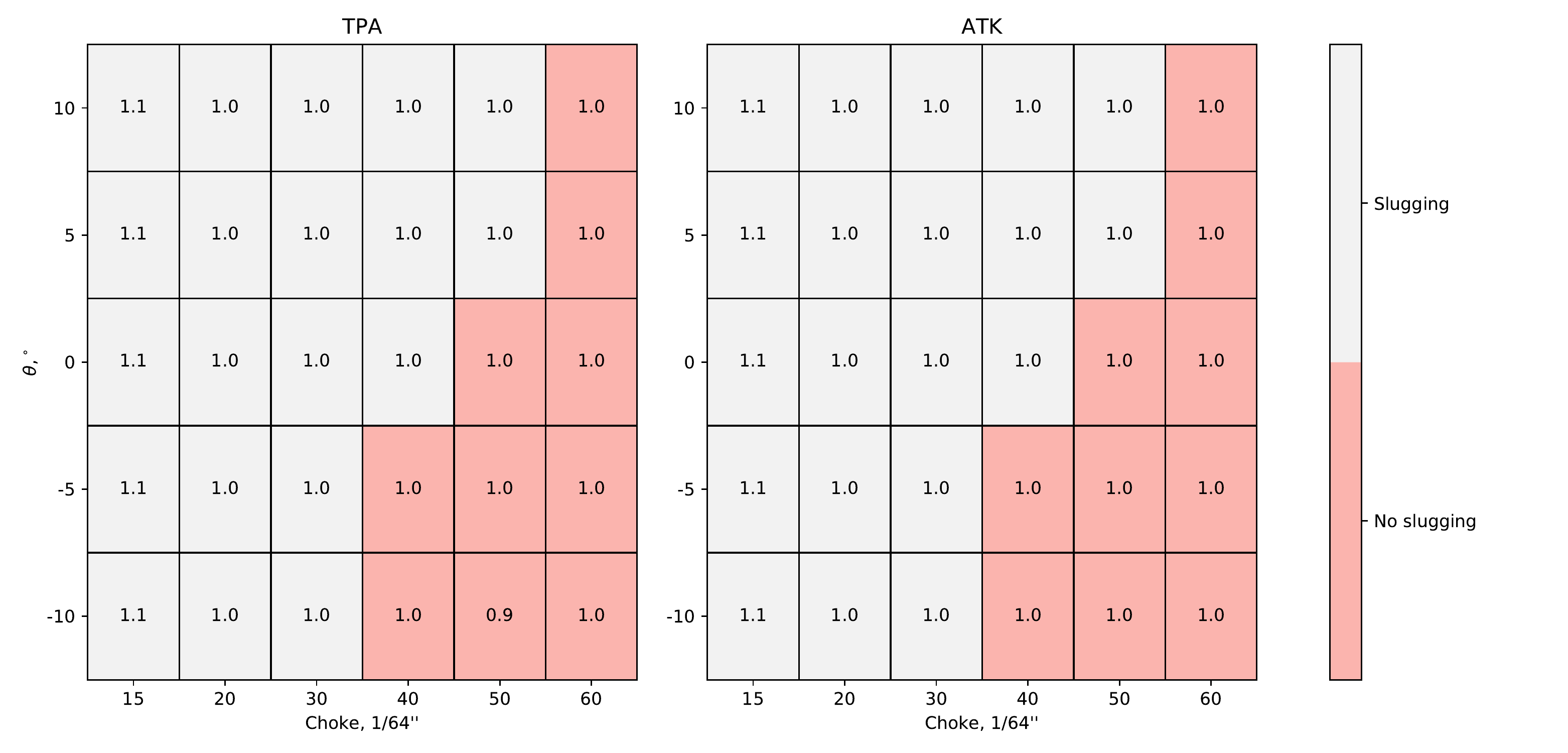}
        \caption{Maximal acceleration for  $\tau = 20$ s.}
        \label{fig:acceleration20}
    \end{figure*}
    The results of computations for the same set of test cases, but with reduced time step $\tau = 20$ s are presented in \tablename~\ref{tab:slug_dt20} and \figurename~\ref{fig:acceleration20}. 
    We can see that the difference in total number of iterations between methods diminishes. 
    This is because as the initial guess at step $t_k$ is the converged solution from previous time step $t_{k-1}$.
    As $\tau$ becomes smaller, the consequent solutions become closer to each other; 
        the initial approximation becomes better, and all solvers need less steps to converge.
    Note that, despite the amount of coupling steps per timestep decreased, the number of timesteps and therefore the total amount of coupling iterations for the whole case increased.
    The maximal acceleration  $1.14$ and $1.09$ for ATK and TPA, respectively, was observed  on the case with $c = 15/64'',\ \theta = -10^\circ$.
    
    \begin{table*}[ht]
        \centering
        \input{tab/tpa150.tex}
        \caption{TPA performance for different values of $\beta$, $\tau = 150\ s$}
        \label{tab:tpa_perf150}
    \end{table*}     
    \tablename\ \ref{tab:tpa_perf150} represents the data for the performance of TPA method on the same set of cases for $\tau = 150\ s$, but at different values of relaxation parameter $\beta$.
    One can clearly see that the amount of iterations is decreasing with increasing $\beta$ . 
    In most cases, either $\beta_{opt} = 1$, or $\beta_{opt} = 0.95$, but the total amount of iterations for $\beta = 1$ is within $3\%$ of the minimal result.
    On the other hand, some deviations from that trend are present (underlined in \tablename\ \ref{tab:tpa_perf150}) for $c =15/64''$. 
    $\beta_{opt}$ is $0.95$ for $\theta = -5^\circ$, $\beta_{opt}$ is $0.75$ for $\theta = 5^\circ$ and result for $\beta = 1$ is $14\%$ and $20\%$ worse than optimal, respectively. 
    From these findings we can suggest that $\beta = 1$ is a good default choice for most of the cases, but one could opt for additional fine tuning to achieve the fastest possible convergence rate under specific combination of model parameters. 
    
    To analyze the performance of the ATK solver and its dependence on the initial relaxation factor $\beta_0$, let us return to the \tablename~\ref{tab:slug_dt150}. 
    A notable finding is that the total number of iterations depends on this parameter much less than it depends on $\beta$ for TPA and FPI solver.
    To provide a quantitative estimate, we consider the relative difference between $N_{avg}$, the number of iterations averaged over $\beta_0$ in given case, and $N_{best}$, minimal iterations required in the case.
    $N_{avg}$ is considered a rough estimate of the expected amount of iterations when $\beta_0$ is chosen completely arbitrarily.
    The maximal relative difference of around $12\%$ between $N_{avg}$ and $N_{best}$ with $c = 15/64''$ and inclinations $0^\circ,\ 5^\circ$. 
    For three other cases it is less than $6\%$, and for five other it is less than $0.6\%$.
    The same relative difference is around $65-72\%$ for TPA, and $61-86\%$ for FPI method, so usage of a completely arbitrary relaxation factor for these methods without any tuning leads to drastic slowdown of convergence.
    Overall, in the slugging test case, ATK provides the same or slightly higher performance than TPA with an additional advantage that it does not require fine tuning of relaxation parameter.
    
    Please note once again that all the numerical experiments for Anderson's acceleration method are done only for a particular case of the algorithm where the next iteration depends only on two previous iterates. 
    Recommendations on the choice of $\beta$ and estimates of performance should be carefully if at all extrapolated on the implementations of Anderson's algorithm using more than two previous iterates.
    It was chosen to use implementation with only two iterates because it requires only the most basic vector operations and the least additional memory. 
    Nevertheless, this implementation provided more than sufficient improvement of performance for our goals. 
    Usage of more previous iterates can further improve the performance of the solver, 
        but then the need to efficiently solve the minimization problem \ref{eq:AAopt} and monitor the conditioning number of the related matrices arises. 
    More details on the multipoint Anderson's method can be found in \cite{WalkerPengAnderson, LottWalkerWoodwardYang}.

%% file: tab/cascading.tex
\begin{tabular}{|l|rrrrrrr|rrr|}
\cline{1-11}
$\beta$ &   0.001 &   0.01 &    0.1 &    0.3 &    0.5 &    0.7 &    1.0 & $\beta_{opt}$ & $N_{best}$ & $S_{max}$ \\
Alg. &         &        &        &        &        &        &        &               &            &           \\
\cline{1-11}
TPA  &  56,088 & 18,810 &  9,672 & 13,971 &  2,274 & 11,910 &  1,095 &         1.000 &      1,095 &       600 \\
ATK  &  59,943 & 59,920 & 65,892 & 67,302 & 75,006 & 75,074 & 73,556 &         0.010 &     59,920 &        11 \\
FPI  & 656,888 &    \multicolumn{6}{c|}{Did not converge}            &         0.001 &    656,888 &       --- \\
\cline{1-11}
\end{tabular}


%% file: tab/slug150_var.tex
\begin{tabular}{|ll|rrr|rrr|rrr|}
\cline{1-11}
     & Alg. & \multicolumn{3}{l|}{FPI} & \multicolumn{3}{l|}{ATK} & \multicolumn{3}{l|}{TPA} \\
     & $\beta$ &    0.1 &    0.5 &    1.0 &   0.1 &   0.5 &   1.0 &    0.1 &   0.5 &   1.0 \\
$c,\ 1/64''$ & $\theta,\ ^\circ$ &        &        &        &       &       &       &        &       &       \\
\cline{1-11}
\multirow{3}{*}{15.0} & -5.0 & 60,123 & 10,680 & 10,353 & 4,465 & 4,210 & 4,793 & 31,373 & 5,998 & 4,924 \\
     & 0.0 & 60,395 &  9,685 &  9,393 & 3,700 & 4,707 & 4,246 & 19,763 & 4,737 & 3,741 \\
     & 5.0 & 60,739 &  9,990 &  8,918 & 3,545 & 4,496 & 4,003 & 29,777 & 6,075 & 4,436 \\
\cline{1-11}
\multirow{3}{*}{40.0} & -5.0 & 62,742 &  9,746 &  3,550 & 3,627 & 3,680 & 3,624 & 30,075 & 5,979 & 3,682 \\
     & 0.0 & 62,591 &  9,727 &  3,453 & 3,584 & 3,608 & 3,601 & 29,749 & 5,988 & 3,628 \\
     & 5.0 & 62,697 &  9,760 &  3,455 & 3,507 & 3,527 & 3,531 & 29,040 & 5,738 & 3,611 \\
\cline{1-11}
\multirow{3}{*}{60.0} & -5.0 & 65,458 & 10,614 &  4,352 & 4,749 & 4,430 & 4,668 & 28,009 & 6,576 & 4,392 \\
     & 0.0 & 63,312 &  9,815 &  3,608 & 3,670 & 3,722 & 3,679 & 27,389 & 6,002 & 3,704 \\
     & 5.0 & 62,300 &  9,680 &  3,514 & 3,620 & 3,587 & 3,612 & 27,494 & 5,819 & 3,630 \\
\cline{1-11}
Avg. &     & 62,261 &  9,966 &  5,621 & 3,829 & 3,996 & 3,973 & 28,074 & 5,879 & 3,972 \\
\cline{1-11}
\end{tabular}

%% file: tab/slug20_var.tex
\begin{tabular}{|ll|rrr|rrr|rrr|}
\cline{1-11}
     & Alg. & \multicolumn{3}{l|}{FPI} & \multicolumn{3}{l|}{ATK} & \multicolumn{3}{l|}{TPA} \\
     & $\beta$ &     0.1 &    0.5 &    1.0 &    0.1 &    0.5 &    1.0 &     0.1 &    0.5 &    1.0 \\
$c,\ 1/64''$ & $\theta,\ ^\circ$ &         &        &        &        &        &        &         &        &        \\
\cline{1-11}
\multirow{3}{*}{15.0} & -5.0 & 405,675 & 62,659 & 14,179 & 12,872 & 12,950 & 12,858 & 129,817 & 26,687 & 13,208 \\
     & 0.0 & 410,835 & 63,855 & 14,188 & 13,173 & 13,243 & 13,284 & 169,768 & 25,840 & 13,164 \\
     & 5.0 & 410,930 & 63,750 & 14,398 & 13,217 & 13,210 & 13,226 & 125,375 & 27,234 & 13,199 \\
\cline{1-11}
\multirow{3}{*}{40.0} & -5.0 & 431,409 & 66,709 & 15,692 & 15,830 & 15,791 & 15,820 & 146,306 & 30,667 & 16,132 \\
     & 0.0 & 429,591 & 66,685 & 15,648 & 15,501 & 15,539 & 15,535 & 155,729 & 30,924 & 16,137 \\
     & 5.0 & 429,429 & 66,969 & 15,739 & 15,568 & 15,637 & 15,576 & 151,272 & 33,076 & 16,123 \\
\cline{1-11}
\multirow{3}{*}{60.0} & -5.0 & 462,578 & 71,354 & 16,489 & 16,787 & 16,730 & 16,734 & 136,651 & 31,470 & 16,421 \\
     & 0.0 & 455,892 & 70,083 & 16,199 & 16,488 & 16,492 & 16,471 & 152,419 & 31,293 & 16,420 \\
     & 5.0 & 456,655 & 70,201 & 16,180 & 16,481 & 16,438 & 16,421 & 155,762 & 32,589 & 16,270 \\
\cline{1-11}
Avg. &     & 432,554 & 66,918 & 15,412 & 15,101 & 15,114 & 15,102 & 147,011 & 29,975 & 15,230 \\
\cline{1-11}
\end{tabular}

%% file: tab/tpa150.tex
\begin{tabular}{|ll|rrr|rrr|rrr|}
\cline{1-11}
    \multicolumn{2}{|l|}{$c,\ 1/64''$}     & \multicolumn{3}{c|}{15.0} & \multicolumn{3}{c|}{40.0} & \multicolumn{3}{c|}{60.0} \\
    \multicolumn{2}{|l|}{$\theta, ^\circ$} &   -5.0 &    0.0 &    5.0 &   -5.0 &    0.0 &    5.0 &   -5.0 &    0.0 &    5.0 \\
\cline{1-11}
\multirow{6}{*}{$\beta$} & 0.10 &  31,373             &  19,763 &  29,777             &  30,075 &  29,749 &  29,040 &  28,009 &  27,389 &  27,494 \\
                         & 0.25 &  13,950             &   9,101 &  12,351             &  12,581 &  12,570 &  12,616 &  12,756 &  12,096 &  11,816 \\
                         & 0.50 &   5,998             &   4,737 &   6,075             &   5,979 &   5,988 &   5,738 &   6,576 &   6,002 &   5,819 \\
                         & 0.75 &   4,357             &   3,848 &   \underline{3,620} &   4,007 &   3,983 &   3,972 &   5,332 &   4,033 &   3,992 \\
                         & 0.95 &   \underline{4,296} &   3,793 &   3,763             &   3,697 &   3,636 &   3,605 &   4,716 &   3,690 &   3,596 \\
                         & 1.00 &   4,924             &   3,741 &   4,436             &   3,682 &   3,628 &   3,611 &   4,392 &   3,704 &   3,630 \\
\cline{1-11}
\end{tabular}

%% file: s5_conclusion.tex
\section{Conclusion}\label{sec:conclusion}
Results of our study indicate that accelerated Anderson's and Aitken's algorithms are handy tools for the solution of fixed-point problems arising, for example, in coupling of systems in multiphysics modelling. The widely used and simple fixed-point iterations algorithm~---~the Picard method~---~has several downsides. One of them is slow convergence. The other one is dependence on a manually tuned relaxation factor. In some dynamic evolution-type problems, there may be no single value of relaxation factor that ensures convergence for the entire duration of the physical process. Specifics of the coupling problem, when it is necessary to iteratively couple existing numerical modules, often impose specific constraints on the solution. Typical constraints are unavailability of analytical Jacobian and computationally expensive target function. This is the case with the well-fracture coupling, which is one of the important problems in e.g. oilfield application, and which was used here for demonstration.
To cope with these constraints, Anderson's and Aitken's methods are suggested. 
Neither of the methods  relies on the usage of the Jacobian, its inverse, or their full-rank approximations and has negligible memory and CPU requirements compared to target function calculation. The target function is evaluated only once per iteration. 
Although it's not guaranteed mathematically that the global convergence properties for these methods would be better than for the Picard's method, in practical applications they demonstrate noticeably better performance. 
In this work, we benchmarked and reconfirmed performance of our implementation of the two-point Anderson's (TPA) and Aitken's (ATK) algorithms using practical examples of well-fracture coupling inspired by oilfield applications. We also investigated how performance of these methods depends on algorithm settings and behavior of the simulated problem and generated recommendations for their efficient usage based on numerical experiments.

In the numerical experiments we considered cases with relatively smooth solutions and cases with intermittent, step-like changes in solution.
ATK tends to be equal or slightly better than TPA in cases with relatively smooth variations of pressures and rates.
At the same, TPA can be superior compared to ATK, by an order of magnitude, in cases with abrupt changes of flow rates, but may require some tuning of relaxation factor.

The data obtained from the numerical experiments generally supports the notion presented in the literature that $\beta = 1$ is a good choice of relaxation factor in Anderson's algorithm.
In the majority of cases, the result was either optimal or sufficiently close to the optimum. 
On the other hand, there are some notable deviations from this trend. In these cases, fine-tuning of the relaxation parameter may result in faster convergence.
The analysis above has shown weak dependence of Aitken's method performance on the initial choice of relaxation factor.

From these findings, we suggest the following practical strategy of usage of the accelerated solvers.
Aitken's algorithm can be used as a default solver for most of the problems. With an arbitrary $\beta_0$, it provides significant acceleration in most of cases. In the worst cases we have seen, it is not slower than the Picard method. For harder problems, e.g., with fast intermittent changes of rate, or small choke, the solution may be further accelerated by application of TPA, which may require fine-tuning of $\beta$. In that case, a good starting point is a value $\beta = 1$.

%% file: s6_acknowledgements.tex
\section*{Acknowledgements}
The authors would like to thank Schlumberger for their support and permission to publish this paper. We would also like to acknowledge A. Chaplygin and P. Spesivtsev for various contributions that helped to complete this paper.

%% file: s7_references.tex
\section*{References}
\bibliography{coupling_refs} 
\bibliographystyle{elsarticle-num}